# Machine Learning based Post Processing Artifact Reduction in HEVC Intra Coding

K.R. Rao, Ninad Gorey

Department of Electrical Engineering, University of Texas at Arlington, USA

**ABSTRACT:**

The lossy compression techniques produce various artifacts like blurring, distortion at block bounders, ringing and contouring effects on outputs especially at low bit rates. To reduce those compression artifacts various Convolutional Neural Network (CNN) based post processing techniques have been experimented over recent years. The latest video coding standard HEVC adopts two post processing filtering operations namely de-blocking filter (DBF) followed by sample adaptive offset (SAO). These operations consumes extra signaling bit and becomes an overhead to network. In this paper we proposed a new Deep learning based algorithm on SAO filtering operation. We designed a variable filter size Sub-layered Deeper CNN (SDCNN) architecture to improve filtering operation and incorporated large stride convolutional, deconvolution layers for further speed up. We also demonstrated that deeper architecture model can effectively be trained with the features learnt in a shallow network using data augmentation and transfer learning based techniques. Experimental results shows that our proposed network outperforms other networks in terms on PSNR and SSIM measurements on widely available benchmark video sequences and also perform an average of 4.1 % bit rate reduction as compared to HEVC baseline.

## 1. INTRODUCTION:

In this era of data explosion, the lossy compression technique such as HEVC is inevitable to save the network bandwidth. Most compression algorithms rely on tiling the images into blocks, applying quantization and sparse transform. Again high compression rates would introduce undesired complex artifacts and compressed images often become blurred due to the loss of high-frequency components. These artifacts decrease the perceptual visual quality and also affects various low-level image processing routines such as edge detection [13], super-resolution [10] and contrast enhancement [2], [19]. Despite the success of video compression techniques, the effective compression artifacts reduction remains an active area of research.

The HEVC standard has included DBF and SAO post processing techniques for artifact reduction. Again these kind of artifact reduction methods can hardly be extended from one compression method to another compression scheme. As a result, a data-driven learning-based methods can be an alternative for better generalization and are considered in recent years. The existing De-blocking algorithms focus on removing blocking and ringing artifacts. To eliminate undesired complex artifacts, the ARCNN [6] model proposed by embedding feature enhancement layers. Kim and Lee [3] proposed super resolution method using very deep residual learning. Kuanar et al [7] proposed a deep learning based in-loop filtering for decoder quality enhancement, Park and Kim [12] proposed a CNN based network to replace the de-blocking filter in HEVC. To remove the undesired noisy features our SDCNN network included feature extraction, de-noising and enhancement layers. But we faced difficulty while training the layers due to sub-optimal initialization. To speed up training we learnt residual image and used high learning rate with adjustable gradient clipping technique [3]. We found that effective learning could be solved by transferring the features learnt in a shallow network to a deeper network and fine-tuning the various data augmentation techniques. To accelerate training we implemented the transfer learning technique on our deep convolutional network [3], [9], [21]. Finally our network effectively suppressed compression artifacts to great extent while retaining edge patterns, sharp details and outperformed previously studied networks in achieving higher bit rate reduction and computational speed.

The remainder of this paper is organized as follows. Section II Proposed Framework, Section III Experiments and Section IV Conclusions.

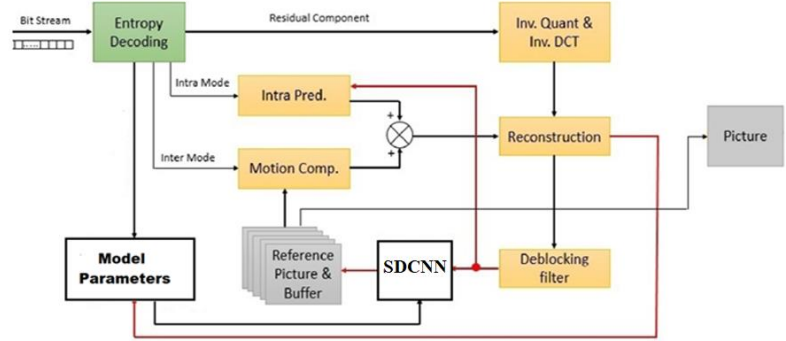

HEVC Decoder Diagram with SAO filter replaced by SDCNN
Figure: 1 HEVC Decoder Diagram

## 2. PROPOSED FRAMEWORK:

In this paper we presented a CNN based network for the compression artifact reduction in HEVC intra coding. Our proposed approach was inspired by the current success of low level computer vision models [3], [5], [6]. Fig. 1 shows our decoder block diagram with newly added SDCNN network.

Our CNN based network takes the compressed image as input from HEVC encoder and output reconstructed residual image. The overall network (Fig. 2) contains six convolutional layers and operated jointly in an end to end framework. The features extracted on first layer was quite noisy. The mapping layer maps above high dimension vector to another high dimensional vector which restores the features. Finally the residual layer aggregated the patch wise representations with strided deconvolution and added input frame to generate final output. The whole network is expressed as:

$F_0(Y) = Y;$
$F_i(Y) = \max(0, W_i * Y + B_i), i \in \{1, 2, 3, 4, 5, 6\};$
$F(Y) = W_6 * F_6(Y) + B_6 + Y$

Where $W_i$, $B_i$ and $F_i$ represents filter, biase, output feature map of the $i_{th}$ layer respectively and '∗' denotes the convolution operation. The $W_i$ contains $n_i$ filters of support $n_{i-1} \times f_i \times f_i$, where $f_i$ was spatial size of a filter, $n_0$ was the number of channels in input image and $n_i$ was the number of filters. Our framework was denoted as "$n_1(f_1) - n_2(f_2) - n_3(f_3) - n_4(f_4) - n_5[f_5] - n_6[f_6] - s[6]$" where f, n, [ ], s represent the filter size, number of filters, deconvolution filter and stride respectively. The network settings are described in Table I and represented as 64(9)-32(3)-32(5)-32(3)-64(3)-1(9)-s. The spatial shrinking and nonlinear mapping layers were chosen with 3×3 filters to reduce overall complexity. We used a strided convolution and deconvolution technique with a stride s of two for network speedup, while preserving the image reconstruction quality. In our experiment we considered Y luminance color frames only and become input to SDCNN. ReLU activation was applied on the filter responses. Our CNN network was designed to learn the residue. The final output frame F(Y) was just a combination of input and residual frame.

## B. Loss function and Model Learning

We calculated the Mean Squared Error (MSE) as a loss function from given the set of ground truth images {Xi} and their corresponding compressed images {Yi}:

$$L(\theta) = \frac{1}{n} \sum_{i=1}^{n} \| F(Y_i, \theta) - X_i \|^2 \quad (1)$$

Where $\Theta = \{W_1 \text{ to } W_6, B_1 \text{ to } B_6\}$, n is number of training samples. Our goal was to learn a model that predicts values $F_i$. Learning end-to-end mapping function F requires estimating the parameters $\Theta$. This was achieved through minimizing loss between the reconstructed residual images $F_i(Y_i; \Theta)$ and the corresponding ground truth frames $X_i$.

TABLE I: THE CONFIGURATION OF SDCNN NETWORK

| Layers | Layer 1 | Layer 2 | Layer 3 | Layer 4 | Layer 5 | Layer 6 |
|---|---|---|---|---|---|---|
| Convolution | Conv1 | Conv2 | Conv3 | Conv4 | Conv5 | Conv6 |
| Filter Size | $f_1$: 9×9 | $f_2$: 3×3 | $f_3$: 5×5 | $f_4$: 3×3 | $f_5$: 3×3 | $f_6$: 9×9 |
| Filters | $n_1$: 64 | $n_2$: 32 | $n_3$: 16 | $n_4$: 32 | $n_5$: 64 | $n_6$: 1 |
| Parameters | 5184 | 18432 | 12800 | 4608 | 18362 | 5164 |
| Total Parameters = 64620 ||||||
| Percentage(%) | 8.11 | 28.51 | 19.8 | 7.12 | 27.91 | 8.51 |

## C. Transfer Model Learning

In practice training entire convolution network from scratch is relatively rare because of insufficient datasets. Instead it is common to pre-train a CNN on a very large dataset and use it either as an initialization or a fixed feature extractor for the task of our interest [4]. To address this issue He et al. [18] derived an initialization method for rectifier nonlinearities and Simonyan et al. [21] used the pre-trained layer features on a shallow network for initialization. We encountered the convergence problem difficulty during our network training and systematically investigated several transfer settings to reuse features learnt in a relatively easier task to initialize a deeper network [15], [8]. But the features learnt in different layered tasks always had a lot in common. These features could be reused during fine tuning and made the convergence faster. So we reused features learnt in a relatively easier task to initialize a deeper network. For our HEVC in-look filtering experiment we defined a base task as $A_{base}$ and two target tasks as $targerBase_i$, $i \in \{1, 2\}$ as shown in Fig. 3. The base network $A_{base}$ was a five layer SDCNN network trained on a large dataset $data_1$, of which frames were compressed using HEVC compression schemes [1], [11] and with a compression quality $Q_1$. For our learning purpose we transfer first few layers of $A_{base}$ to different target tasks $targetBase_1$ and $targetBase_2$. In our case the target layers were six layers with different quantization parameter (QP) values. We followed two transfer learning models 1) shallow-deeper and 2) high-low model for our comparative experiments.

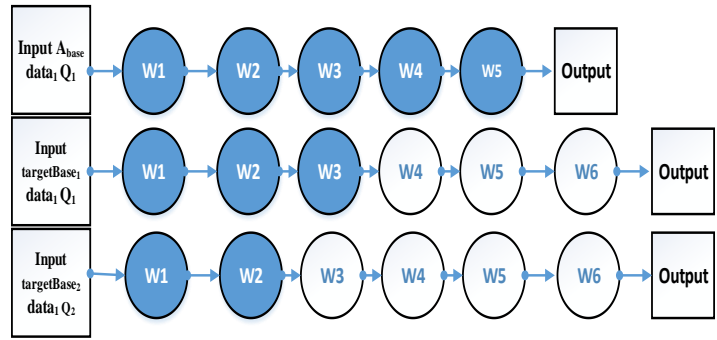

Fig. 3. Transfer Learning Settings.

While learning shallow-deeper model we transferred first 3 layers of $A_{base}$ to six layered network $targetBase_1$. Then we randomly initialized its remaining layers and trained all layers towards same dataset $data_1$. Again it was observed that images with low compression quality contain more complex artifacts. In our high-low model learning we used features learnt from high compression quality frames as starting point and then learnt more complicated features in deep network. The first 2 layer of $targetBase_2$ were copied from $A_{base}$ and trained on compressed frames with a lower compression quality $Q_2$

## 3. EXPERIMENTS:

For our training experiment we randomly selected 400 input images from BSDS500 dataset [22] and used HEVC standard video sequences for testing. The training and testing images had no overlap to demonstrate the generalizability. To use the dataset more efficiently we adopted three data augmentation techniques 1) horizontal flipping 2) Scaling by a factor of 0.75, 0.5 and 0.25 3) Rotation by degrees of 90, and 180. Then our augmented training set is 2×4×3= 24 times of original one.

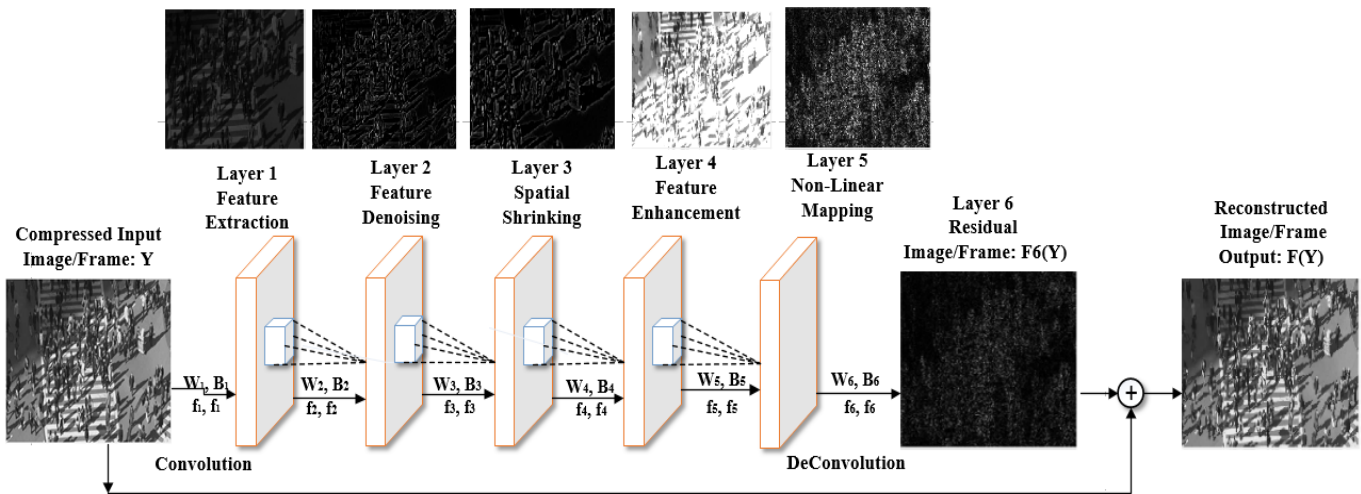

Fig. 2. Framework of Multi-layered Deeper Convolutional Neural Network (SDCNN)

We trained our SDCNN network on Tensorflow [10] framework using NVIDIA K40 GPU machine. Training set images were decomposed into 32 × 32 sub-images $X = \{X_i\}$, $i \in (1 \ldots n)$. We used HEVC HM 16.9 software [11] for our experiment. Each sub-image $X_i$ was compressed by HEVC intra coding at 4 different quantization parameters: QP = 22, 27, 32, 37. Then compressed samples Y = {$Y_i$}, i ∈ (1 ... n) were generated from training samples. The sub-images were extracted from ground truth images with a stride of 10. Thus the augmented 400×32=12800 training images provided 2,421,376 training samples. Separate network were trained for each QP. The loss function L (Θ) was computed by comparing de-convoluted output image of 32 × 32 pixels and ground-truth sub-image $X_i$. The loss was minimized using stochastic gradient descent (SGD) with back propagation. In training phase we used a smaller learning rate $0.5×10^{-4}$ in the last layers for better convergence and comparably larger one $0.5×10^{-3}$ in remaining layers. The momentum parameter was set to 0.99 and weight decay to 0.001. We trained over 6420 iterations with batch size 128. The trained files along with different QP parameters compiled and integrated on locally installed HM software. For our testing purpose we turn off SAO filter from configuration file and performed an end to end flow

A. **Comparison with HEVC**

We tested our network with our standard video sequences classes A, B, C and D and their corresponding BD-bitrate results are presented on Table II. We compared our SDCNN network post processing results against the HEVC DBF and SAO outputs. To evaluate coding efficiency we used the BD-rate measure [17] on luminance and chrominance channels independently. It was observed that our model achieved significant bit rate reduction in all sequences listed in Table II. For the luminance (Y), a high of 7.2% BD-rate was achieved on the BlowingBubbles sequences and on average 4.1% BD-rate was achieved on all the sequences

TABLE II: AVERAGE BD-RATE OF MDCNN COMPARED TO HEVC

| Class | Sequence | Frame Count | BD-rate Y% | U% | V% |
|---|---|---|---|---|---|
| Class A Resolution: (2560×1600) | PeopleOnStreet | 150 | -4.9 | -5.7 | 5.7 |
| | Traffic | 150 | -5.4 | -3.5 | -4.1 |
| | SteamLocomotive | 300 | -1.9 | -0.5 | -0.3 |
| | Nebuta | 300 | -0.9 | -4.6 | -4.1 |
| Class B Resolution: (1920×1080) | BQTerrace | 600 | -2.4 | -1.7 | -1.3 |
| | ParkScene | 500 | -4.3 | -3.4 | -2.6 |
| | BasketballDrive | 500 | -2.6 | -3.5 | -5.3 |
| | Kimono | 240 | -2.4 | -1.7 | -1.3 |
| Class C Resolution: (832 × 480) | BQMall | 600 | -4.9 | -5.1 | -5.0 |
| | PartyScene | 500 | -3.4 | -3.9 | -4.4 |
| | BasketballDrill | 500 | -6.9 | -5.7 | -6.8 |
| | RaceHorses2 | 300 | -4.1 | -6.2 | -10.5 |
| Class D Resolution: (416 × 240) | BlowingBubbles | 500 | **-7.2** | -8.5 | **-11.0** |
| | RaceHorses1 | 300 | -4.7 | -8.3 | -7.6 |
| | BasketballPass | 500 | -3.8 | -4.3 | -6.5 |
| | BQSquare | 600 | -3.4 | -3.9 | -4.4 |
| Class Summary | | Class A | -3.4 | -3.6 | -3.5 |
| | | Class B | -3.5 | -3.0 | -3.4 |
| | | Class C | -4.7 | -5.3 | -6.7 |
| | | Class D | -5.6 | -5.3 | -5.1 |
| Overall | | All | **-4.1** | **-4.5** | **-5.3** |

B. **Comparison with other listed models:**

We compared our SDCNN network with HEVC Baseline [1], MDCNN [7] and ARCNN [6]. The comparative results of BD-PSNR and SSIM for test sequences were shown in Table III. As a whole SDCNN outperformed other models based on the evaluation metrics by a large margin. This indicates that SDCNN can produce frames with less compression artifacts. Again we compared the Computational Complexity of different models in terms of GPU decoding time on Class D sequences with resolution $416 \times 240$. All methods were implemented in Caffe framework and 16GB Tesla K40 GPU machine. Experimental result showed that our proposed SDCNN was faster (0.56 sec/frame) than the ARCNN (0.84 sec/frame), VRCNN (0.98 sec/frame) and slower than HEVC (0.26 sec/frame) in spite of our deeper architecture. It was also observed that the decoding time did not meet the real time application requirement and needs further optimization.

C. **Various SDCNN transfer learning model comparisons**

On Table IV we have shown the different transfer learning settings followed during our SDCNN model training. The base network ($A_{base}$) was a five-layer network 9-3-7-3-9 trained on traffic video sequence at compression quality Q22 and BSDS500 [22] dataset. The network parameters were initialized from a Gaussian distribution with zero mean and 0.002 standard deviation. Then we experimented our six layered network 9-3-5-3-3-9 using 'shallow to deep' transfer model and 'high to low' transfer quality techniques. Results in Fig. 4 show that the transferred features from a five-layer network enabled us to train a six-layer network successfully. It was observed that the "shallow to deeper" network converged faster and achieved better performance than learning from scratch random initialization and He et al.'s initialization [18].

TABLE IV: EXPERIMENTAL SETTING OF TRANSFER LEARNINGS

| Transfer Strategy | Short Form | Network Structure | Initialization Strategy |
|---|---|---|---|
| Base network | base Q22 | 9-3-7-3-9 | Gaussian(0, 0.002) |
|  | base Q32 | 9-3-5-3-3-9 | Gaussian(0, 0.002) |
| Shallow to deep | base Q22 | 9-3-7-3-9 | Gaussian(0, 0.002) |
|  | Transfer Deeper | 9-3-5-3-3-9 | 1,2,3 layers of base Q22 |
|  | He et.al [18] | 9-3-5-3-3-9 | He et.al [18] |
| High to Low | base Q22 | 9-3-7-3-9 | Gaussian(0, 0.002) |
|  | Transfer 2 Layers | 9-3-7-3-9 | 1,2 layers of base Q22 |
|  | Transfer 3 Layers | 9-3-7-3-9 | 1,2,3 layers of base Q32 |

In our 'high to low quality' transfer learning excrement we trained our six layered network with quality factor Q22 (low), Q32 (high) and initialized with the Gaussian distribution. Subsequently we performed two experiments for our training: 1) Transferred first 2 layers of base Q32 learnt parameters to base Q22 (Transfer-2 Layers) and 2) Transferred first 3 layers of base Q32 learnt parameters to base Q22 (Transfer 3 Layers). It was observed from Fig. 5 that Transfer-2 Layers outperformed Transfer-3 Layers but could not converge fully.

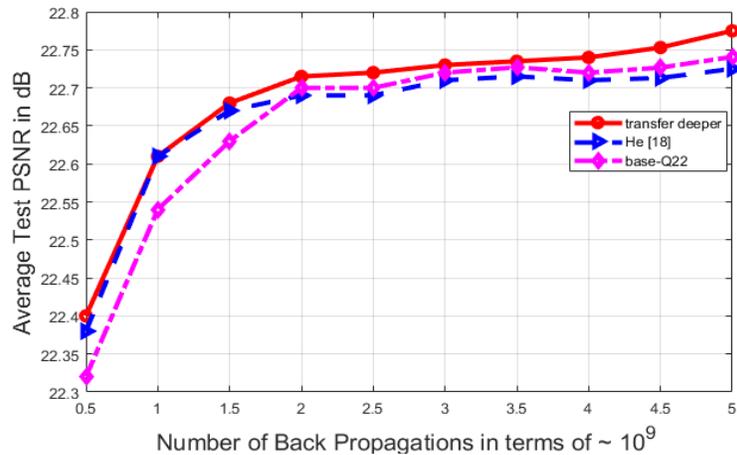
Fig. 4. Transfer Learning Shallow to Deep network model experiment

We compared the visual quality of different architectures with our proposed SDCNN network. For our experiment we used 'PeopleOnStreet' video frames and the reconstructed images are shown on Fig. 6. It is observed our SDCNN was able to eliminate compression artifacts to a great extent as compared to VRCNN and HEVC base line.

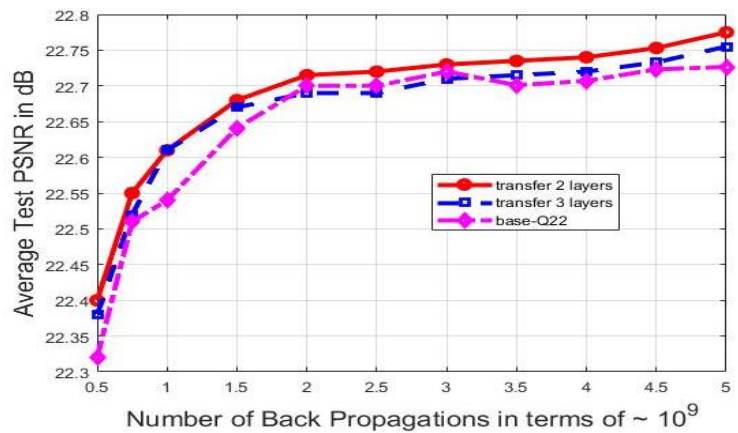
Fig. 5. Transfer Learning High to Low Quality model experiment

## 4. CONCLUSIONS:

In this paper we presented a deep convolutional neural network based HEVC in loop SAO filtering for Intra (I) frame video analysis. Our proposed network SDCNN outperformed the previously studied ARCNN and HEVC base line in terms of achieving higher bit rate reduction and capable of learning the nonlinearity between input and reconstructed images. Although our deep learning based network shows a possible alternative to in-loop filters, more careful transfer training strategies are important for the real-time usages. In the future work we would like to include broader dataset categories to obtain more powerful and generic features. The recurrent neural network (RNN) is widely used learning based algorithm for video analysis [14], [16]. We would also like to explore the Gated Recurrent Unit (GRU) LSTM network [20] and bidirectional LSTM [23], for the inter frame (P and B) predictions and investigate further.

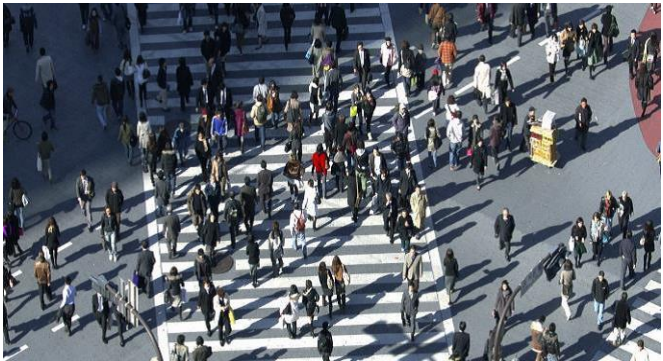
PeopleOnStreet 5<sup>th</sup> frame 2560x1600

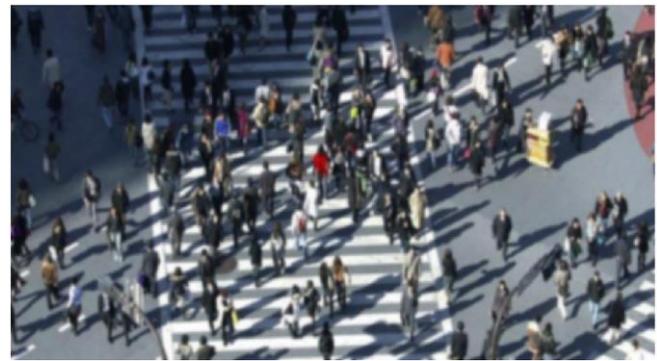
HEVC baseline, PSNR: 29.31 dB

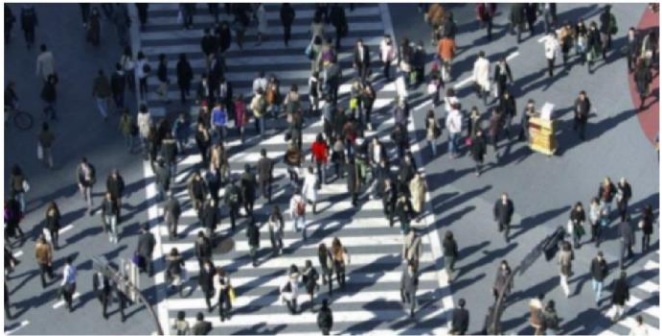
MDCNN, PSNR 30.45 dB

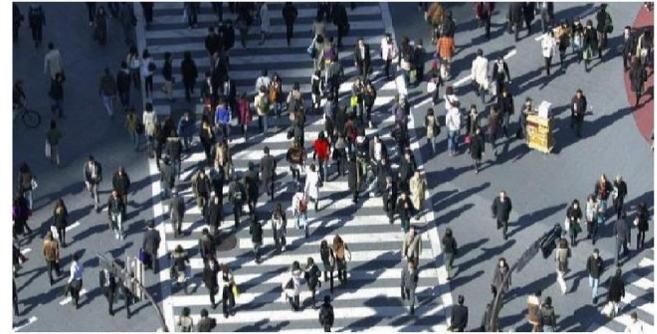
Our CNN Architecture PSNR 33. 01 dB

Fig. 6 Artifacts visualization of compressed frames by different architectures